\documentclass[twocolumn,twoside,slac]{revtex4}
\usepackage{graphicx}
\usepackage{fancyhdr}
\renewcommand{\headrulewidth}{0pt}
\renewcommand{\footrulewidth}{0pt}

\begin{document}

\title{Managing Dynamic User Communities in a Grid of Autonomous Resources}

%

\author{R. Alfieri, R. Cecchini, V. Ciaschini, L. dell'Agnello\footnote{Corresponding author}, A. Gianoli,
F. Spataro}
\affiliation{INFN, Italy}

\author{F. Bonnassieux}
\affiliation{CNRS, France}

\author{P. Broadfoot, G. Lowe}
\affiliation{University of Oxford, United Kingdom}

\author{L. Cornwall, J. Jensen, D. Kelsey}
\affiliation{CCCCLRC/Rutherford Appleton Laboratory, United
Kingdom}

\author{\'A. Frohner}
\affiliation{CERN}

\author{D.L. Groep, W. Som de Cerff, M. Steenbakkers, G. Venekamp}
\affiliation{NIKHEF, The Netherlands}

\author{D. Kouril}
\affiliation{CESNET, Czech Republic}

\author{A. McNab}
\affiliation{University of Manchester, United Kingdom}

\author{O. Mulmo}
\affiliation{KTH, Sweden}

\author{M. Silander,  J. Hahkala}
\affiliation{HIP, Finland}

\author{K. L\H{o}rentey}
\affiliation{ELTE, Hungary}

\begin{abstract}
One of the fundamental concepts in Grid computing is the creation
of Virtual Organizations (VO's): a set of resource consumers and
providers that join forces to solve a common problem. Typical
examples of Virtual Organizations include collaborations formed
around the Large Hadron Collider (LHC) experiments. To date, Grid
computing has been applied on a relatively small scale, linking
dozens of users to a dozen resources, and management of these VO's
was a largely manual operation. With the advance of large
collaboration, linking more than 10000 users with a 1000 sites in
150 counties, a comprehensive, automated management system is
required. It should be simple enough not to deter users, while at
the same time ensuring local site autonomy. The VO Management
Service (VOMS), developed by the EU DataGrid and DataTAG
projects\cite{edg,edt}, is a secured system for managing
authorization for users and resources in virtual organizations. It
extends the existing Grid Security Infrastructure\cite{gsi}
architecture with embedded VO affiliation assertions that can be
independently verified by all VO members and resource providers.
Within the EU DataGrid project, Grid services for job submission,
file- and database access are being equipped with fine- grained
authorization systems that take VO membership into account. These
also give resource owners the ability to ensure site security and
enforce local access policies. This paper will describe the EU
DataGrid security architecture, the VO membership service and the
local site enforcement mechanisms \textbf{Local Centre
Authorization Service} (LCAS), \textbf{Local Credential Mapping
Service}(LCMAPS) and the Java Trust and Authorization Manager.
\end{abstract}

\maketitle

\thispagestyle{fancy}


\section{INTRODUCTION}
The security infrastructure of the European Datagrid (EDG) is
based on Public Key Infrastructure (PKI)\cite{PKI}. There are a
number of independent entities, known as the national Certificate
Authorities (CA), that are responsible for identifying the users
of the grid and assigning them with authentication credentials
(i.e. X.509 certificates); for example, in this project, there are
18 CA's.

Following a process of harmonization of Certificate Policy (CP)
and Certificate Practice Statements (CPS), all these CA's mutually
trust each other and are trusted by all resources participating in
the EDG test-bed.

On the other hand, user authorization is a much more complex
mechanism and constitutes one of the most challenging issues in
Grid computing.

Since resources and users are not typically co-located, it is not
feasible to make Authorization decisions for grid users on local
(i.e. on the resource) site basis only. Moreover users have
normally direct administrative deals only with their own local
site (or a few sites) and with the collaborations they work in,
but not, generally, with all the entities forming a grid.

Throughout this paper, we will refer to the following terminology:

\begin{itemize}

\item {\bfseries Virtual Organization} ({\bfseries VO}): abstract
entity grouping Users, Institutions and Resources (if any) in a
same administrative domain \cite{fos,fos2};

\item {\bfseries Resource Provider} ({\bfseries RP}): facility
offering resources (e.g. CPU, network, storage) to other parties
(e.g. VO's), according to specific ``Memorandum of
Understanding''.

\end{itemize}

In the LHC Computing Grid (LCG)\cite{lcg} framework, the LHC
experiments collaborations are good examples of VO's, while the
Tiers are the RP's.

From the authorization point of view, a grid is established by
enforcing agreements between RP's and VO's, where, in general,
resource access is controlled by both parties with different
roles. More specifically, the general information regarding the
relationship of the user with his VO (groups he belongs to, roles
he is allowed to cover \footnote{For the definition of group and
role, please see \cite{far}.} and capabilities\footnote{A
capability is intended as a free text string to be interpreted by
the local site for special processing purposes.} he should present
to RP's for special processing needs) should be managed by the VO
itself. The RP's evaluate locally this information, granted by the
VO to the user, taking into account the local policies and the
agreements with the VO.

As result of the authorization evaluation process, the RP should
eventually grant access to a certain set of resources (CPU,
storage etc.) in a controlled way (e.g. by some kind of extended
Access Control Lists) and provide mapping from the grid
credentials (i.e. X.509 certificates) onto local ones (e.g. Unix
credentials on computing elements).

Since, in our opinion, the present mechanisms (e.g. grid-mapfile,
VO LDAP servers) are not scalable (as discussed earlier in this
Section), it is unfeasible to use them in a production grid, with
a potential very large number of users (e.g. exceeding thousands
of people); hence we have developed a new set of tools to cope
with all these aspects.

In this paper we briefly describe the authorization requirements
of a grid, focusing on the framework of the DataGrid and DataTAG
(EDT) Projects, and illustrate the architecture of the new
services we have developed, the \textbf{Virtual Organization
Membership Service} (VOMS) to manage authorization information in
VO scope, the  LCAS to handle the actual authorization decision at
the local site, the LCMAPS to map the grid credentials to local
ones, \textbf{TrustManager} to provide GSI compatible
authentication in Java and the \textbf{Authorization Manager} for
coarse grained access control in Java services.


\subsection{Authorization Requirements}

Given the large potential number of users and resources in a grid,
the Authorization infrastructure for users needs to be centralized
at VO level to satisfy the requirement of scalability. Therefore,
as stated before, Authorization is based on policies written by
VO's and their agreements with RP's, that, in turn, enforce local
authorization.

In this VO-centric vision, the first condition for a user to
access the grid is to be member of a VO, but in general a user may
be member of any number of VO's and, for this reason, his
membership should be considered as a ``reserved'' information
(i.e. its membership in a VO must not be relevant to other VO's).

A VO can form a complex, hierarchical structure with groups and
subgroups in order to clearly divide its users according to their
tasks: in general, we can represent the VO structure with a Direct
Acyclic Graph (DAG), where the groups are the vertices of the
graph and the subgroup-group relationships are the oriented edges.

For scalability reasons, it should be possible to delegate the
management of each group.

A user can be a member of any number of these groups and, both at
VO and group level, may be characterized by any number of roles
and capabilities; moreover roles and capabilities may be granted
to the user indefinitely or on a scheduled time basis (e.g. a
certain user of the CMS collaboration is granted administrative
role only when he is ``on shift'') or on a periodic basis (e.g.
normal users have access to resources only during working hours).
The membership in a subgroup implies the membership in all
ancestor groups up to the root (i.e. the VO itself).

The enforcement of these VO-managed policy attributes (group
memberships, roles, capabilities) at local level must reflect the
agreements between the VO and the RP's. However it should be
possible for RP's to override the permissions granted by VO's
(e.g. to ban unwanted users). As a consequence, to permit
traceability at user level (and not only at VO level) users must
present their credential to RP's along with their authorization
info.

As side requirements, we note that authorization servers should
not be a single point of failure (this is a particularly critical
issue for VO authorization servers) and that all the Authorization
communications should be trusted, secured and reserved.

\subsection{Authorization status in EDG}\label{testbed1}

The Authentication and Authorization methods adopted by the EDG
are based on the Globus Toolkit's Grid Security Infrastructure
(GSI) \cite{globus,gsi} and on compatible solutions for Java.

In EDG, as originally in Globus, to access the Grid, the user
first creates a proxy certificate (via grid-proxy-init procedure)
that is then sent to the requested resources in order to access
them.

\begin{figure}[htbp]
\begin{center}
\includegraphics[width=0.45\textwidth]{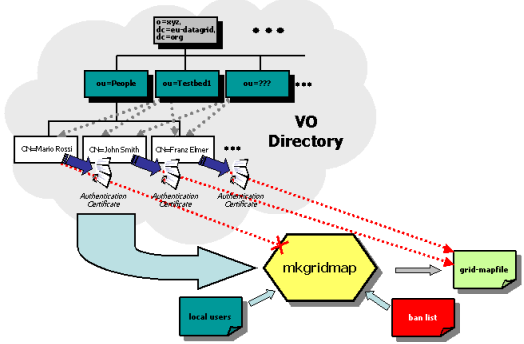}
\caption{\label{figure:VOLDAP}VO LDAP authorization mechanism}
\end{center}
\end{figure}

In EDG Test-bed 1 each VO maintains authorization information in a
LDAP server. The RP's, periodically (e.g. daily) querying the LDAP
servers, generate a list of VO users (in case banning unwanted
entries or allowing non-VO users on their local resources) and map
them to local credentials (the so-called ``grid-mapfil'') granting
users the Authorization to access local resources. A tool,
\emph{mkgridmap}, to generate the list is also available (see
Figure~\ref{figure:VOLDAP}).

In the current implementation, authorization is boolean since
neither subgroups nor roles or capabilities are supported; hence a
differentiation among users is only manageable at the local sites
and can only reflect local policies.


The main missing features of this architecture are flexibility and
scalability. No roles, subgroups memberships and any other user
peculiarity are supported. Moreover, the use of a RP-based
database (i.e. the grid-mapfile), periodically updated, hardly
scales in a production environment with a large number of users,
each, potentially, with his groups, roles and capabilities,
whereas in the test-bed the users situation is almost static, and
user policy is very simple.

The solution is to let users present the authorization data as
they try to access the local resources (i.e. shifting from pull to
push model); on the other hand we suspect that LDAP protocol is
not the best choice to sustain the burden of a potentially high
number of complex queries.

\section{The VOMS system}
The Virtual Organization Membership Service (VOMS) has been
developed in the framework of EDG and EDT collaborations to solve
the current LDAP VO servers limitations (see paragraph
\ref{testbed1}). In fact, the purpose of VOMS is to grant
authorization data to users at VO level.

VOMS provides support for group membership, "forced" groups (i.e.
for negative permissions), roles (e.g. admin, student, etc.) and
capabilities (free form string).

\begin{figure}[htbp]
\begin{center}
\includegraphics[width=0.45\textwidth]{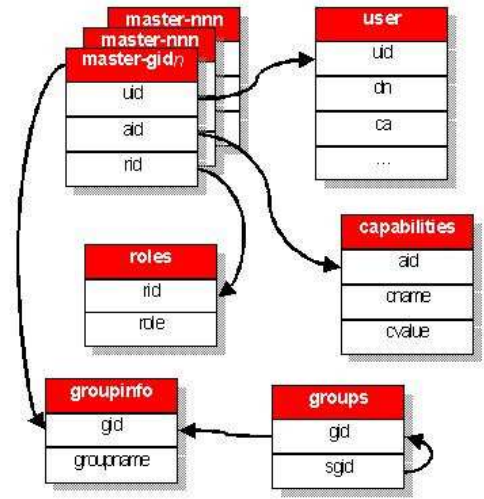}
\caption{\label{figure:groups}The VOMS database structure}
\end{center}
\end{figure}

The server is essentially a front-end to a Relational Database
Management System (RDBMS) (in the present implementation, the
database used is mysql\cite{mysql}), where all the information
about users is kept (see Figure~\ref{figure:groups}).

\begin{figure}[htbp]
\begin{center}
\includegraphics[width=0.45\textwidth]{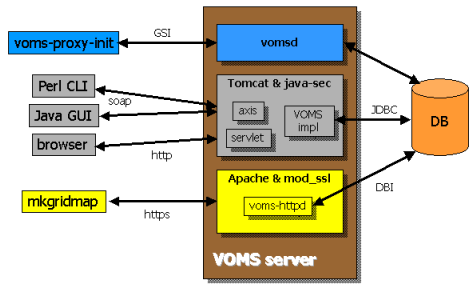}
\caption{\label{figure:VOMS}The VOMS system}
\end{center}
\end{figure}

\noindent The VOMS System is composed by the following parts (see
figure \ref{figure:VOMS}):

\begin{itemize}

\item {\bfseries User Server}: receives requests from a client and
returns information about the user.

\item {\bfseries User Client}: contacts the server presenting a
user's certificate and obtains a list of groups, roles and
capabilities of the user. All client-server communications are
secured and authenticated.

\item {\bfseries Administration Client}: used by the VO
administrators (adding users, creating new groups, changing roles,
etc.)

\item {\bfseries Administration Server}: accepts the requests from
the clients and updates the Database.

\end{itemize}

\subsection{Operations}

\subsubsection{User part}

For continuity reasons with the present situation, we have added a
command ({\tt voms-proxy-init}) to replace {\tt grid-proxy-init}.
This new command produces a user's proxy certificate -- like {\tt
grid-proxy-init} -- but with the difference that it contains the
user authorization info from the VOMS server(s). This info is
returned in a structure containing also the credentials both of
the user and of the VOMS server and the time validity. All these
data are signed by the VOMS server itself. We call this structure
a ``Pseudo-Certificate'' (a new release with Attribute
Certificates \cite{hou,far} replacing the ``Pseudo-Certificate''
is in progress).

\begin{figure}[htbp]
\begin{center}
\includegraphics[width=60mm]{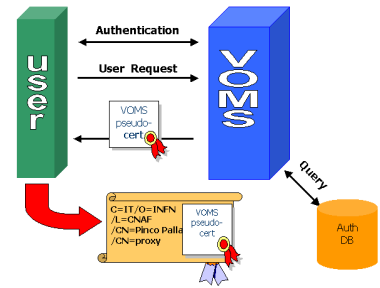}
\end{center}
\caption{\label{figure:VOMSoperations}VOMS Operations}
\end{figure}

\noindent In more detail (see figure \ref{figure:VOMSoperations}):

\begin{enumerate}

\item The user and the VOMS server mutually authenticate using
their certificates (via standard Globus API);

\item The user sends signed request to VOMS Server;

\item The VOMS Server checks correctness\footnote{The server,
after checking the user's signature, verifies the congruity of the
requested info.} of user's request;

\item The VOMS Server sends back to the user the required info
(signed by itself) in a structured form ("Pseudo-Certificate");

\item The user checks the validity of the info received;

\item The user eventually repeats process for other VOMS's;

\item The user creates the proxy certificates containing all the
info received from the VOMS Server into a (non critical)
extension;

\item The user may add user-supplied authentication info (kerberos
tickets, etc…).

\end{enumerate}

In order to process this authorization information, the
\emph{Gatekeeper}, in addition to normal certificate checking, has
to extract the additional information embedded in the proxy (the
Pseudo-Certificate). This can be easily done with an appropriate
LCAS plug-in \cite{wp4}. However, as the VOMS info are  included
in a non critical extension of the certificate, this can be used
even by ``VOMS-unaware'' \emph{Gatekeepers}, thus maintaining
compatibility with previous releases.

For the transition phase, we have also developed an enhanced
version of \emph{mkgridmap} (\emph{mkgridmap++}) to allow RP's,
with the old \emph{Gatekeeper} installation, to query the VOMS
server instead of the LDAP server. We control the access to VOMS
allowing only authenticated users via https.

The Java counterpart of LCAS/LCMAPS, the \emph{Authorization
Manager} \cite{d76} is also capable of parsing and checking the
VOMS Pseudo-Certificate and utilise its attributes in the
authorization process. To ease the transition it is backward
compatible and it is also able to use a grid-mapfile.

\subsubsection{Administration\\}

The Administration server supports the SOAP protocol for
connections, so that it can be easily converted into an OGSA
service. It consists of five sets of routines, grouped into
services: the {\bfseries Core}, which provides the basic
functionality for the clients; the {\bfseries Admin}, which
provides the methods to administrate the VOMS database; the
{\bfseries History}, which provides the logging and auditing
functionality (the database scheme provides full audit records for
every changes); {\bfseries Request}, which provides an integrated
request handling mechanism for new users and for other changes;
{\bfseries Compatibility}, which provides a simple access to the
user list for the \emph{mkgridmap} utility.

Two administrative interfaces (web and command line) are
available.

\subsection{Security Considerations}

The authentication to the VOMS server makes use of the standard
GSI security controls on the user's certificate before granting
rights: it must be signed by a ``trusted'' CA, be valid and not
revoked.

Compromising the VOMS server could allow granting credentials with
access rights for any service itself would be not enough to grant
illegal, indiscriminate access. However, it should kept in mind
that the authorization data must be inserted in a user proxy
certificate (i.e. countersigned by the user himself). Large scale
vulnerabilities are denial of service attacks (e.g. to prevent VO
users to get their authorization credentials).

The main security issue about proxy certificates\cite{proxy} is
the lack of a revocation mechanism; on the other hand these
certificates have short lifetimes (12 hours, typically).

\subsection{Related Works}

In this paragraph, we will briefly compare the VOMS system with
some analogous systems, namely the ``Privilege and Role Management
Infrastructure Standards Validation'' (PERMIS), Akenti and the
``Community Authorization Server'' (CAS).

\subsubsection{VOMS vs. PERMIS}

PERMIS\cite{permis}, implementing an RBAC (Role Based Access
Control) mechanism, has been considered as an alternative to VOMS.

PERMIS has two modes of operation, push and pull. With push, the
user sends his attribute certificates to PERMIS; with pull, PERMIS
can be configured with any number of LDAP repositories, and it
will search all of them for attributes of the user.

This second approach is clearly neither VOMS-oriented nor
scalable.

Moreover VOMS, distributing the AC's to the users themselves,
allows a much greater flexibility. For example, with VOMS a user
who is a member of several groups and holds several roles can
actually choose how much information about himself he may want to
present to a site. It is also possible to obtain and present at
the same time information on more VO's, a useful characteristic in
case of collaborations between VO's.

The second major difference is the policy engine, where Permis is
really powerful, because it can take a properly formatted policy
file and make decisions based on the content of the file and the
AC's it receives. On the contrary, VOMS does not focus on this
problem, and it leave the interpretation of the AC's to other
components (i.e. to local sites, namely to LCAS).

In conclusion VOMS and Permis are complementary: VOMS as a AC
issuer, and Permis (slightly modified in its AC gathering) as an
policy engine.

\subsubsection{VOMS vs. CAS}

CAS\cite{cas} has been developed by the Globus team to solve the
same problem tackled by VOMS in EDG.

There are, indeed, two major differences between CAS and VOMS.

The first is that CAS does not issue AC's, but whole new proxy
certificates with the CAS server Distinguish Name as the subject;
the authorization information is included in an extension.

As a consequence, when a service receives this certificate, it
cannot effectively decide who the owner is without inspecting the
extension. This means that existing services, in Globus-based
grids, would need to be modified to use a CAS certificate; on the
contrary using VOMS, since it adds the AC's in a non-critical
extension of a standard proxy certificate, does not require this
kind of modification to the services.

The second major difference is in the fact that CAS does not
record groups or roles, but only permissions. This means that the
ultimate decision about what happens in a farm is removed from the
farm administrator and put in the hands of the CAS administrator,
thus breaking one of the fundamental rules of the grid: the farm
administrator has total control about what happens on his
machines.

\subsubsection{VOMS vs. Akenti}

Akenti\cite{akenti} is an AC-based authorization system.

There are three major differences between Akenti and VOMS.

The first is that Akenti does not use true AC's since their
definition and description do not conform the standard (at present
nor VOMS uses standard AC's, but this will be changed in the next
production release).

The second is that Akenti is targeted on authorizing accesses on
web resources, and particularly web-sites.  This means that it is
completely unfeasible to use it for other needs, for example in a
VO.

The third is that Akenti does not link identities with groups or
roles, but with permissions. This is done on the resource side,
not removing the control from the resource itself, like CAS does;
on the other hand, not having an intermediary like VOMS (or even
CAS) will surely lead to fragmentation and inconsistencies between
the permissions.

\begin{figure*}[t]
\centering
\includegraphics[width=90mm, angle=90]{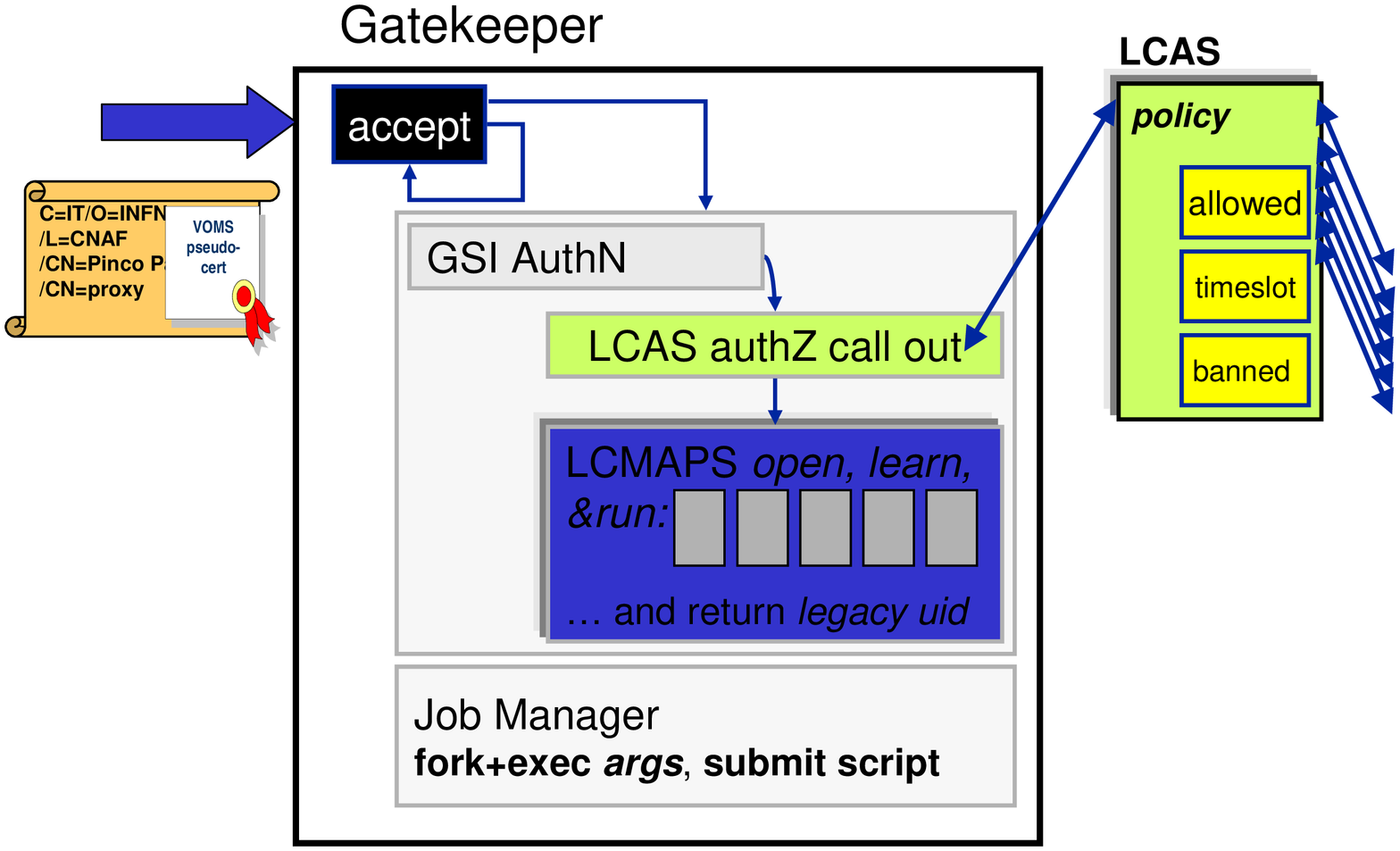}
\caption{\label{figure:gatekeeper}The (EDG modified) gatekeeper}
\end{figure*}

\section{Local Authorization Services}
At the resource level, two new services have been introduced to
process the authorization data provided by VOMS:
\begin{enumerate}
\item
  For native execution environments, like UNIX, the Local Centre Authorization
  Service (LCAS) and the Local Credential Mapping Service (LCMAPS) replace the
  existing grid-mapfile mechanism, since the traditional system as shipped with
  the \emph{Globus} ``gatekeeper'' allowed credential mapping based only on user
  identity. The gatekeeper has been modified to support these new systems (see figure \ref{figure:gatekeeper}).
\item
  The new Java based web services (e.g. Replica Manager, Spitfire) are based on the GSI compatible authenticaten and coarse grained
authorization routines.
\end{enumerate}

\subsection{LCAS}

The LCAS system provides a pluggable framework for (possibly
centralized) site authorization. LCAS is called from within the
\emph{gatekeeper}. Based on identity, authorization data, and the
complete job specification, access to can be granted or denied.
Several plug-in modules are shipped by default with the system,
amongst them modules to support site-specific blacklisting of
users and wall-clock time constraints on job submission. RP's can
develop and subsequently include locally developed modules.

A plug-in for VOMS (to process Authorization data) has been
developed.

\subsection{LCMAPS}

Traditional native execution environments like UNIX, have no ready
means to enforce the specific set of rights represented in the
Authorization data used in the Grid environment. For such
environments, those rights have to be \emph{mapped} to credential
mechanisms supported by these native environments, like \emph{user
id}s and \emph{group id}s. The mapping created should result in
the rights and limitation as expressed on the Grid be enforced
when running jobs on a particular system. The mapping is thus
based on user identity, VO affiliation and, necessarily,
site-local policies. LCMAPS is, like LCAS, a pluggable framework
system. It supports both standard UNIX (\emph{uid,gid})-pair
accounts, either predefined for individual users, or allocated
on-the-fly from a pool of generic ``leased'' accounts via the
Gridmapdir mechanisms\cite{gridmapdir}. These mappings can be
enforced both in the local process as well as in central user
directories based on LDAP. Further plug-ins support the
acquisition of local Kerberos5 credentials and AFS tokens.

\subsection{Java Security}


The TrustManager is certificate validator, authentication
subsystem for Java services: it supports the X.509 certificates
(and CRL's) and GSI style proxy certificates, making possible to
(mutually) authenticate the client-server connections.


In the EDG Java security package the authorization is implemented
as role based authorization. Currently the authorization is done
in the server end and the server authorizes the user, but there
are plans to do mutual authorization where also the client end
checks that the server end is authorized to perform the service or
to save the data.

\section{Future Developments}

Future developments will include use of Attribute Certificates for
all the Authorization process, replica mechanisms for the RDBMS
containing the Authorization data, and more sophisticate time
validity for the VOMS certificates.

\begin{acknowledgments}
The authors wish to thank the EU and our national funding agencies
for their support.
\end{acknowledgments}


\end{document}